\documentclass[aps,preprint,showpacs,superscriptaddress]{revtex4}%
\usepackage{amsfonts}
\usepackage{amsmath}
\usepackage{amssymb}
\usepackage{graphicx}%
\begin{document}
\preprint{HEP/123-qed}
\title{Manipulation and readout of spin states of a single-molecule magnet by a
spin-polarized current}
\author{Hai-Bin Xue}
\email{xuehaibin@tyut.edu.cn}
\affiliation{Key Laboratory of Interface Science and Engineering in Advanced Materials,
Ministry of Education, Taiyuan University of Technology, Taiyuan, Shanxi
030024, China}
\author{Jiu-Qing Liang}
\affiliation{Institute of Theoretical Physics, Shanxi University, Taiyuan, Shanxi 030006, China}
\author{Wu-Ming Liu}
\affiliation{Beijing National Laboratory for Condensed Matter Physics, Institute of
Physics, Chinese Academy of Sciences, Beijing, 100190, China.}

\begin{abstract}
Single-molecule memory device based on a single-molecule magnet (SMM) is one
of the ultimate goals of semiconductor nanofabrication technologies. Here, we
study how to manipulate and readout the SMM's two spin-state of stored
information that characterized by the maximum and minimum average value of the
$Z$-component of the total spin of the SMM and the conduction-electron,
which are recognized as the information bits \textquotedblleft$1$" and \textquotedblleft$0$".
We demonstrate that the switching time depends on both the sequential tunneling gap $\varepsilon_{se}$
and the spin-selection-rule allowed transition-energy $\varepsilon_{trans}$,
which can be tuned by the gate voltage. In particular, when the external bias
voltage is turned off, in the cases of the unoccupied and doubly-occupied
ground eigenstates, the time derivative of the transport current can be used
to read out the SMM's two spin-state of stored information. Moreover, the
tunneling strength of and the asymmetry of the SMM-electrode coupling have a
strong influence on the switching time, but that have a slight influence on
the readout time that being on the order of nanoseconds. Our results suggest a
SMM-based memory device, and provide fundamental insight into the electrical
controllable manipulation and readout of the SMM's two spin-state of stored information.

\end{abstract}
\date{\today}
\maketitle

\section{INTRODUCTION}

The realization of memory device based on single-molecule magnet (SMM) is
currently one of the ultimate goals of semiconductor nanofabrication
technologies. In particular, the SMM memory devices provide the final
density-limit of magnetic storage media
\cite{GuoFS2018,SessoliR,GoodwinCAP,GuoFS,Delgado}. Consequently, the
single-molecule magnet (SMM), as a smallest magnetic bistable system
\cite{GoodwinCAP,GuoFS,Heersche,Jo,Zyazin,Kahle,Burzuri,Sun,MisiornyM}, has
been suggested as a promising candidate for the information storage and
processing due to its chemical tunability and scalability \cite{Bogani}. A
fundamental step towards realization of the SMM-based memory devices is the
manipulation and detection of spin states of the SMM. In the nanoscale
devices, a local magnetic field is required to manipulate the spin degree of
freedom. In resent years, using electric fields instead of magnetic fields to
manipulate the spin states of the nanomagnets (i.e., the SMM, magnetic atoms
on surfaces and magnetic molecules) have become the subject of intensive
experimental
\cite{GodfrinEx01,GodfrinEx02,ThieleEx,VincentEx,KomedaEx,NattererEx,YanEx,WarnerEx,LothEx}
and theoretical
\cite{Timm,Misiorny07,Misiorny08,Misiorny09,Lu,Misiorny13,ZhangZZ,Karlewski,DelgadoPRL,Liljeroth,
Hirjibehedin,Gartlanda,Bode,DelgadoPRB,ABComas,HymasK19,Wrzessniewski20,HymasK20}
investigations due to the electric fields being easily focused and shielded
within a small volume.

The electric-field manipulation of spin states of a SMM can be realized by
transferring the spin angular momentum of conduction electrons to the SMM via
the the spin transfer torque. For example, the switching of spin states in the
SMM can be controlled with an external spin-polarized current through the SMM,
which is coupled with two electrodes either one \cite{Timm,ZhangZZ} or both of
them being ferromagnetic \cite{Misiorny07,Misiorny08,Misiorny09,Misiorny13,Wrzessniewski20},
or applied by a spin bias voltage \cite{Lu}. The giant-spin orientation of the SMM can be
determined by the direction of spin-polarized current \cite{Timm,Misiorny13,ZhangZZ}.
From the SMM-based memory device point of
view, there are at least two aspects of the bias-voltage-driven switching of
the SMM's spin-state that should be considered. Firstly, what are the
essential factors that governed the switching time of the SMM's two spin-state of stored information?
Secondly, how to read out the SMM's two spin-state of stored information? The
previous works are mainly focused on the first aspect, for example, the
influences of the bias voltage, the spin polarizations of the source and drain
electrodes, and the SMM's internal level structure on the switching processes
of the SMM's spin states. However, the dependence of the switching time and
the readout time on the SMM's ground eigenstate have not yet been revealed. In
particular, the readout scheme of and the readout time of the SMM's
two spin-state of stored information are still some open issues and require further studies.

In this work, we study the influences of the SMM's ground eigenstates on the
switching time and the readout time, and discuss how to manipulate and read
out the SMM's two spin-state of stored information. It is numerically
demonstrated that the minimum switching time, which is strongly affected by
the tunneling strength of and the asymmetry of SMM-electrode coupling, can be
observed when the energy eigenvalues of the SMM's two singly-occupied
eigenstates approach that of the unoccupied or doubly-occupied ones. In particular,
in the cases of the unoccupied and doubly-occupied ground eigenstates,
the effective readout scheme of the SMM's two spin-state of stored information
is proposed by turning off the external bias voltage.
The corresponding readout time of the SMM's two spin-state of stored information is of
the order of nanoseconds. These results suggest the feasibility of an
individual SMM as single-molecule memory device.

\section{MODEL AND FORMALISM}

\subsection{Hamiltonian of a single-molecule magnet coupled to two electrodes}

The single-molecule memory device consists of a SMM weakly coupled to the two
ferromagnetic leads, as shown schematically in Fig. 1. Here, we consider a
prototypical SMM Mn$_{12}$ (Mn$_{12}$-acetate) \cite{Heersche,Jo,Kahle,Sun},
which has total spin $S=10$ and an anisotropy barrier of $6$ meV. The SMM
Hamiltonian can be described by
\begin{equation}
H_{\text{SMM}}=(\varepsilon_{\text{LUMO}}-eV_{g})\hat{n}+\frac{U}{2}\hat
{n}(\hat{n}-1)-J\,\vec{s}\cdot\vec{S}-K_{2}(S_{z})^{2}, \label{model}%
\end{equation}
Here, the first and second terms describe the lowest unoccupied non-degenerate
molecular orbital (LUMO) with on-site energy $\varepsilon_{\text{LUMO}}$,
which can be tuned by a spin-independent gate voltage $V_{g}$. $\hat{n}\equiv
d_{\uparrow}^{\dag}d_{\uparrow}+d_{\downarrow}^{\dag}d_{\downarrow}$ with
$d_{\sigma}^{\dag}$ ($d_{\sigma}$) being the creation (annihilation) operator
of an electron with spin $\sigma$ for the LUMO. $U$ is the on-site Coulomb
repulsion between two electrons in the LUMO. The third term denotes the
exchange coupling between the conduction-electron spin $\vec{s}\equiv
\sum_{\sigma\sigma^{\prime}}d_{\sigma}^{\dag}\left(  \vec{\sigma}%
_{\sigma\sigma^{\prime}}\right)  d_{\sigma^{\prime}}$\ and the giant spin
$\vec{S}$ of the SMM, where $\vec{\sigma}\equiv$ $(\sigma_{x},\sigma
_{y},\sigma_{z})$ is the vector of Pauli matrices. The forth term is the
anisotropy energy of the SMM, whose easy-axis is $Z$-axis ($K_{2}>0$), thus,
the spin is quantized along the $Z$-axis.

The spin relaxation in the two electrodes is assumed to be sufficiently fast,
thus, their electron distributions can be modeled as non-interacting Fermi
gases and the Hamiltonian reads
\begin{equation}
H_{\text{electrodes}}=\sum_{\alpha\mathbf{k}s}\varepsilon_{\alpha\mathbf{k}%
s}a_{\alpha\mathbf{k}s}^{\dag}a_{\alpha\mathbf{k}s}, \label{Leads}%
\end{equation}
where $a_{\alpha\mathbf{k}\sigma}^{\dag}$ ($a_{\alpha\mathbf{k}\sigma}$) is
the creation (annihilation) operator of an electron with energy $\varepsilon
_{\alpha\mathbf{k}\sigma}$, momentum $\mathbf{k}$ and spin $s$ in $\alpha$
($\alpha=L,R$) electrode, with the index $s=+$ ($-$) denoting the majority
(minority) spin states with the density of states $g_{\alpha}^{s}$. The
electrode polarization is characterized by the orientation of the polarization
vector $\mathbf{P}_{\alpha}$ and its magnitude is defined as $P_{\alpha
}=(g_{\alpha}^{+}-g_{\alpha}^{-})/(g_{\alpha}^{+}+g_{\alpha}^{-})$. Here, the
polarization vectors $\mathbf{P}_{L}$ (left electrode) and $\mathbf{P}_{R}$
(right electrode) are parallel or antiparallel to the spin quantization
$Z$-axis. In the parallel case, spin-up $\uparrow$ and spin-down $\downarrow$
are defined to be the majority spin and minority spin of the ferromagnetic
electrode, respectively. The tunneling Hamiltonian between the SMM and the two
electrodes is thus described by
\begin{equation}
H_{\text{tunneling}}=\sum_{\alpha\mathbf{k}\sigma}\left(  t_{\alpha
\mathbf{k}\sigma}a_{\alpha\mathbf{k}\sigma}^{\dag}d_{\sigma}+\text{H.c.}%
\right)  , \label{tunneling}%
\end{equation}
In the ferromagnetic electrode case, the spin-dependent tunneling rate is
given as $\Gamma_{\alpha}^{\sigma}=2\pi|t_{\alpha}|^{2}g_{\alpha}^{\sigma
}=(1+\sigma P_{\alpha})\Gamma_{\alpha}/2$ and $\Gamma_{\alpha}=\Gamma_{\alpha
}^{\uparrow}+\Gamma_{\alpha}^{\downarrow}$, where the tunneling amplitudes
$t_{\alpha}$ and the density of the state $g_{\alpha}^{\sigma}$ are assumed to
be independent of wave vector and energy. While in the normal-metal electrode
case, $P_{\alpha}=0$ and then $\Gamma_{\alpha}^{\uparrow}=\Gamma_{\alpha
}^{\downarrow}=$ $\Gamma_{\alpha}/2$. In the case of the polarization vector
$\mathbf{P}_{\alpha}$ being antiparallel to the spin quantization $Z$ axis,
the electrode-$\alpha$ spin polarization rate $P_{\alpha}$ takes a negative value.

\subsection{Energy eigenstates of the isolated SMM}

In terms of the four electron states of the LUMO $\left\vert 0,0\right\rangle
_{\text{LUMO}}$, $\left\vert \uparrow,0\right\rangle _{\text{LUMO}}$,
$\left\vert 0,\downarrow\right\rangle _{\text{LUMO}}$, $\left\vert
\uparrow,\downarrow\right\rangle _{\text{LUMO}}$, and the spin states of the
SMM $\left\vert m\right\rangle _{\text{GS}}$ ($m=-S$, $-S+1$, $\cdots$, $S-1$,
$S$), the unoccupied and doubly-occupied eigenstates
can be expressed as
\begin{equation}
\left\vert 0,m\right\rangle =\left\vert 0,0\right\rangle _{\text{LUMO}%
}\left\vert m\right\rangle _{\text{GS}},m=-S,-S+1,\cdots,S-1,S, \label{zero}%
\end{equation}
and%
\begin{equation}
\left\vert 2,m\right\rangle =\left\vert \uparrow,\downarrow\right\rangle
_{\text{LUMO}}\left\vert m\right\rangle _{\text{GS}},m=-S,-S+1,\cdots,S-1,S,
\label{two}%
\end{equation}
respectively. In the singly-occupied case, the eigenstates of isolated SMM
have the following two branches \cite{Timm}
\begin{equation}
\left\vert 1,m\right\rangle ^{\pm}=a_{m}^{\pm}\left\vert \uparrow
,0\right\rangle _{\text{LUMO}}\left\vert m-\frac{1}{2}\right\rangle
_{\text{GS}}+b_{m}^{\pm}\left\vert 0,\downarrow\right\rangle _{\text{LUMO}%
}\left\vert m+\frac{1}{2}\right\rangle _{\text{GS}}, \label{one}%
\end{equation}
with%
\[
a_{m}^{\pm}=\frac{J\sqrt{S\left(  S+1\right)  -m^{2}+1/4}}{2\sqrt{\Delta
E\left(  m\right)  }\sqrt{2\Delta E\left(  m\right)  \mp\left(  2K_{2}%
-J\right)  m}},
\]%
\[
b_{m}^{\pm}=\mp\frac{\sqrt{2\Delta E\left(  m\right)  \mp\left(
2K_{2}-J\right)  m}}{2\sqrt{\Delta E\left(  m\right)  }}.
\]
where $\Delta E\left(  m\right)  =\left[  K_{2}\left(  K_{2}-J\right)
m^{2}+\left(  J/4\right)  ^{2}\left(  2S+1\right)  ^{2}\right]  ^{1/2}$ and
$m=-S+1/2$, $-S+3/2$, $\cdots$, $S-3/2$, $S-1/2$ that being the magnetic
quantum-number of the total spin of the SMM and the conduction-electron. While in the $m=\pm\left(  S+1/2\right)  $
cases, the two singly-occupied branches are $\left\vert 1,S+1/2\right\rangle
=\left\vert \uparrow,0\right\rangle _{\text{LUMO}}\left\vert S\right\rangle
_{\text{GS}}$ for $m=S+1/2$, and $\left\vert 1,-S-1/2\right\rangle =\left\vert
0,\downarrow\right\rangle _{\text{LUMO}}\left\vert -S\right\rangle
_{\text{GS}}$ for $-S-1/2$. The corresponding energy eigenvalues are given by%
\begin{equation}
\varepsilon\left(  0,m\right)  =-K_{2}m^{2}, \label{energy0}%
\end{equation}%
\begin{equation}
\varepsilon\left(  2,m\right)  =2\left(  \varepsilon_{\text{LUMO}}%
-eV_{g}\right)  +U-K_{2}m^{2}, \label{energy2}%
\end{equation}%
\begin{equation}
\varepsilon^{\pm}\left(  1,m\right)  =\varepsilon_{\text{LUMO}}-eV_{g}%
+\frac{J}{4}-K_{2}\left(  m^{2}+\frac{1}{4}\right)  \pm\Delta E\left(
m\right)  . \label{energy1}%
\end{equation}
Here, for the states $\left\vert 1,\pm\left(  S+1/2\right)  \right\rangle $,
the upper (lower) sign applies if $K_{2}-J/2$ is positive (negative). It is
found from Eqs. (\ref{energy2}) and (\ref{energy1}) that the energy
eigenvalues of the eigenstates $\left\vert 1,m\right\rangle ^{\pm}$ and
$\left\vert 2,m\right\rangle $ can be tuned by an applied gate voltage $V_{g}$.

\subsection{Quantum master equation and dynamics of the density matrix}

Due to the SMM-electrode coupling is sufficiently weak, the sequential
tunneling dominates the transitions. The electron tunneling processes are well
described by quantum master equation of a reduced density matrix in a set of
basis vectors of the SMM's eigenstates. Under the Born approximation and
Markovian approximation, the second-order quantum master equation for the
reduced density matrix is given by\cite{LiQE}
\begin{equation}
\dot{\rho}\left(  t\right)  =-i\mathcal{L}\rho\left(  t\right)  -\frac{1}%
{2}\mathcal{R}\rho\left(  t\right)  , \label{Master1}%
\end{equation}
with%
\begin{equation}
\mathcal{R}\rho\left(  t\right)  =%
{\displaystyle\sum\limits_{\sigma=\uparrow,\downarrow}}
\left[  d_{\sigma}^{\dagger}A_{\sigma}^{\left(  -\right)  }\rho\left(
t\right)  +\rho\left(  t\right)  A_{\sigma}^{\left(  +\right)  }d_{\sigma
}^{\dagger}-A_{\sigma}^{\left(  -\right)  }\rho\left(  t\right)  d_{\sigma
}^{\dagger}-d_{\sigma}^{\dagger}\rho\left(  t\right)  A_{\sigma}^{\left(
+\right)  }\right]  +H.c., \label{Master2}%
\end{equation}
where $A_{\sigma}^{\left(  \pm\right)  }=\sum_{\alpha=L,R}A_{\alpha\sigma
}^{\left(  \pm\right)  }$, $A_{\alpha\sigma}^{\left(  \pm\right)  }%
=\Gamma_{\alpha}^{\sigma}n_{\alpha}^{\pm}\left(  -\mathcal{L}\right)
d_{\sigma}$, $n_{\alpha}^{+}=f_{\alpha},n_{\alpha}^{-}=1-f_{\alpha}$, and
$f_{\alpha}$ is the Fermi function of the electrode $\alpha$. The Liouvillian
superoperator $\mathcal{L}$ is defined as $\mathcal{L}\left(  \cdots\right)
=\left[  H_{\text{SMM}},\left(  \cdots\right)  \right]  $. Throughout this
work, we set $e\equiv\hbar=1$. The differential equation (\ref{Master1}) can
be solved by the fourth order Runge-Kutta method. Here, the average value of
the $Z$-component of the total spin of the SMM and the conduction-electron is defined as $\left\langle S_{z}%
\right\rangle =%
{\displaystyle\sum\nolimits_{n,m}}
mP_{\left\vert n,m\right\rangle }$, where $P_{\left\vert n,m\right\rangle }$
is the occupation probability of the eigenstate $\left\vert n,m\right\rangle
$. The maximum value $\left\langle S_{z}\right\rangle _{\max}$ corresponds to
the SMM's spin being parallel to the $Z$-axis, whereas the minimum one
$\left\langle S_{z}\right\rangle _{\min}$ corresponds to that being
antiparallel to the $Z$-axis. Consequently, the SMM's two spin-state of the
$\left\langle S_{z}\right\rangle _{\max}$ and $\left\langle S_{z}\right\rangle
_{\min}$ are evaluated, respectively, as the two states of the binary code,
namely, the spin-up ($1$) and spin-down ($0$).

\section{RESULTS AND DISCUSSIONS}

We now study the manipulation and readout of the SMM's two spin-state of the
$\left\langle S_{z}\right\rangle _{\max}$ and the $\left\langle S_{z}%
\right\rangle _{\min}$, and discuss the dependence of the switching time and
the readout time on the SMM's ground eigenstate and the tunneling strength of
and the asymmetry of SMM-electrode coupling. The typical parameters of a SMM
Mn$_{12}$ are taken as: $S=10$, $U=25$, $J=0.2$, $K_{2}=0.06$ and $k_{B}%
T=0.4$, where the energy unit is meV. In addition, the bias voltage is assumed
to be symmetrically entirely dropped at the SMM-electrode tunnel junctions,
i.e., $\mu_{L}=-\mu_{R}=V_{b}/2$. Here, the initial spin-state is chosen as a
mixed states consisting of the SMM's two ground eigenstates with the same
occupation probabilities, i.e., $P_{\left\vert n,-S\right\rangle }$=$P_{\left\vert n,S\right\rangle }$=$1/2$ ($n=0,2$) or
$P_{\left\vert 1,-S-1/2\right\rangle }$=$P_{\left\vert 1,S+1/2\right\rangle }$=$1/2$. Thus, the
average value of the SMM's initial spin-state $\left\langle S_{z}\right\rangle
_{\text{init}}=0$.

\subsection{Electron tunneling associated with the SMM's spin-state switching}

The spin angular momentum of the conduction electrons can be transferred to
the SMM in the electron tunneling processes due to the exchange coupling
between electron-spin and the SMM. The electron tunneling channels, which
change the quantum number $m$ of the SMM by $\pm1$, can be expressed as
follows:%
\begin{equation}
\left\{
\begin{array}
[c]{c}%
\left\vert 0,m\right\rangle \overset{\text{into: }\uparrow}{\longrightarrow
}\left\vert \uparrow,0\right\rangle _{\text{LUMO}}\left\vert m\right\rangle
_{\text{GS}}\overset{\text{relax}}{\longrightarrow}\sum_{\nu=\pm}a_{m+\frac
{1}{2}}^{\nu}\left\vert 1,m+\frac{1}{2}\right\rangle ^{\nu}\overset{\text{out:
}\downarrow}{\longrightarrow}\sum_{\nu=\pm}a_{m+\frac{1}{2}}^{\nu}%
b_{m+\frac{1}{2}}^{\nu}\left\vert 0,m+1\right\rangle ,\\
\left\vert 1,m\right\rangle ^{\pm}\overset{\text{into: }\uparrow
}{\longrightarrow}b_{m}^{\pm}\left\vert 2,m+\frac{1}{2}\right\rangle
\overset{\text{out: }\downarrow}{\longrightarrow}b_{m}^{\pm}\left\vert
\uparrow,0\right\rangle _{\text{LUMO}}\left\vert m+\frac{1}{2}\right\rangle
_{\text{GS}}\overset{\text{relax}}{\longrightarrow}b_{m}^{\pm}\sum_{\nu=\pm
}a_{m+1}^{\nu}\left\vert 1,m+1\right\rangle ^{\nu},\\
\left\vert 2,m\right\rangle \overset{\text{out: }\downarrow}{\longrightarrow
}\left\vert \uparrow,0\right\rangle _{\text{LUMO}}\left\vert m\right\rangle
_{\text{GS}}\overset{\text{relax}}{\longrightarrow}\sum_{\nu=\pm}a_{m+\frac
{1}{2}}^{\nu}\left\vert 1,m+\frac{1}{2}\right\rangle ^{\nu}\overset
{\text{into: }\uparrow}{\longrightarrow}\sum_{\nu=\pm}a_{m+\frac{1}{2}}^{\nu
}b_{m+\frac{1}{2}}^{\nu}\left\vert 2,m+1\right\rangle ,
\end{array}
\right.  \label{mintomax}%
\end{equation}%
\begin{equation}
\left\{
\begin{array}
[c]{c}%
\left\vert 0,m\right\rangle \overset{\text{into: }\downarrow}{\longrightarrow
}\left\vert 0,\downarrow\right\rangle _{\text{LUMO}}\left\vert m\right\rangle
_{\text{GS}}\overset{\text{relax}}{\longrightarrow}\sum_{\nu=\pm}b_{m-\frac
{1}{2}}^{\nu}\left\vert 1,m-\frac{1}{2}\right\rangle ^{\nu}\overset{\text{out:
}\uparrow}{\longrightarrow}\sum_{\nu=\pm}b_{m-\frac{1}{2}}^{\nu}a_{m-\frac
{1}{2}}^{\nu}\left\vert 0,m-1\right\rangle ,\\
\left\vert 1,m\right\rangle ^{\pm}\overset{\text{into: }\downarrow
}{\longrightarrow}a_{m}^{\pm}\left\vert 2,m-\frac{1}{2}\right\rangle
\overset{\text{out: }\uparrow}{\longrightarrow}a_{m}^{\pm}\left\vert
0,\downarrow\right\rangle _{\text{LUMO}}\left\vert m-\frac{1}{2}\right\rangle
_{\text{GS}}\overset{\text{relax}}{\longrightarrow}a_{m}^{\pm}\sum_{\nu=\pm
}b_{m-1}^{\nu}\left\vert 1,m-1\right\rangle ^{\nu},\\
\left\vert 2,m\right\rangle \overset{\text{out: }\uparrow}{\longrightarrow
}\left\vert 0,\downarrow\right\rangle _{\text{LUMO}}\left\vert m\right\rangle
_{\text{GS}}\overset{\text{relax}}{\longrightarrow}\sum_{\nu=\pm}b_{m-\frac
{1}{2}}^{\nu}\left\vert 1,m-\frac{1}{2}\right\rangle ^{\nu}\overset
{\text{into: }\downarrow}{\longrightarrow}\sum_{\nu=\pm}b_{m-\frac{1}{2}}%
^{\nu}a_{m-\frac{1}{2}}^{\nu}\left\vert 2,m-1\right\rangle ,
\end{array}
\right.  \label{maxtomin}%
\end{equation}
where the notations \textquotedblleft into: $\sigma$\textquotedblright\ and
\textquotedblleft out: $\sigma$\textquotedblright\ denote the spin $\sigma$
electron tunneling into and out the SMM, respectively. While the notations
\textquotedblleft relax\textquotedblright\ describes the electron state of the
SMM $\left\vert i,j\right\rangle _{\text{LUMO}}\left\vert m\right\rangle
_{\text{GS}}$ relaxing to the corresponding SMM's eigenstates.

The effective switching from the initial spin-state of the $\left\langle
S_{z}\right\rangle _{\text{init}}$ to that of the $\left\langle S_{z}%
\right\rangle _{\max}$ can be realized when the number of spin-up electrons
tunneling into the SMM from the source electrode is larger than that of
spin-down electrons (i.e., $P_{L}>0$), and the number of spin-down electrons
tunneling out the SMM and into the drain electrode is not less than that of
spin-up electrons (i.e., $P_{R}\leq0$), see Fig. 2(a) and the dashed, dash
dotted, and short dotted lines in Figs. 2(b) and 2(c). While the switching
from the initial spin-state of the $\left\langle S_{z}\right\rangle
_{\text{init}}$ to that of the $\left\langle S_{z}\right\rangle _{\min}$ can
take place when $P_{L}\leq0$ and $P_{R}>0$, i.e., reversing the direction of
the bias voltage, see the solid, dotted, and short dashed lines in Figs. 2(b)
and 2(c). Consequently, the information can be stored or written onto the
SMM's spin-state of the $\left\langle S_{z}\right\rangle _{\max}$ or
the $\left\langle S_{z}\right\rangle _{\min}$ by manipulating the direction of
the external bias voltage. Moreover, with the increase of the polarization
magnitude of the source (drain) electrode $P_{L}$ ($P_{R}$), the electron
tunneling processes associated with the SMM's spin state switching is
accelerated, leading to the decreasing of the corresponding switching time,
see Figs 2(b) and 2(c). In the following discussion, we consider a SMM weakly
coupled to two ferromagnetic leads with the antiparallel
high-spin-polarization configuration, and set $P_{L}=-P_{R}=0.9$.

\subsection{Dependence of the switching time on the SMM's ground eigenstate}

In the SMM system, it is clear from Eqs. (\ref{energy0})-(\ref{energy1}) that
the ground eigenstates of the SMM can be tuned by the applied gate voltage.
For given parameters of the SMM Mn$_{12}$, the minimum energy eigenvalue of
the doubly-occupied eigenstate $\varepsilon_{\left\vert 2,S\right\rangle }$
decreases faster than that of the singly-occupied eigenstate $\varepsilon
_{\left\vert 1,S+1/2\right\rangle }$, whereas the minimum energy eigenvalue of the unoccupied
eigenstate $\varepsilon_{\left\vert 0,S\right\rangle }$ are independent of
the applied gate voltage, see Fig. 3. Here, we focus on the region of
energy-level crossing between the singly-occupied and doubly-occupied ground
eigenstates, and that of energy-level crossing between the singly-occupied and
unoccupied ground eigenstates, see the region enclosed by the circle with
dotted line, and study the influences of the LUMO's energy $\varepsilon
_{\text{LUMO}}-eV_{g}$ and the bias voltage on the switching time.

In the first region of the energy-level crossing, with increasing the negative
value of the $\varepsilon_{\text{LUMO}}-V_{g}$, which corresponds to the
transition of the ground eigenstates from the doubly-occupied to
singly-occupied eigenstates, the switching time first decreases and then
increases, see Figs 4(c) and 4(d). While in the second region of the
energy-level crossing, with increasing the value of the $\varepsilon
_{\text{LUMO}}-V_{g}$ from negative values to positive values, which
corresponds to the transition of the ground eigenstates from the
singly-occupied to unoccupied eigenstates, the switching time first also
decreases and then increases, see Figs 4(a) and 4(b).

In fact, for the fixed bias voltage and spin polarizations of the source and
drain electrodes, the switching time depends on both the sequential tunneling
gap $\varepsilon_{se}$ \cite{Aghassi17}, which is the energy difference
between the ground eigenstate of charge $N$ and the first excited eigenstate
of charge $N-1$, and the spin-selection-rule allowed transition-energy
$\varepsilon_{trans}$, which is the difference between the energy eigenstates
with the electron-number $n$ and $n\pm1$ satisfying the spin selection rule.
Consequently, with increasing the value of the $\varepsilon_{\text{LUMO}%
}-V_{g}$ from negative values to positive values, the transition of the
sequential tunneling gap $\varepsilon_{se}$ from $\left\vert \varepsilon
_{\left\vert 1,\pm\left(  S+1/2\right)  \right\rangle }-\varepsilon
_{\left\vert 2,\pm S\right\rangle }\right\vert $ to $\left\vert \varepsilon
_{\left\vert 1,\pm\left(  S+1/2\right)  \right\rangle }-\varepsilon
_{\left\vert 0,\pm S\right\rangle }\right\vert $, and the magnitudes of the
sequential tunneling gap $\varepsilon_{se}$ and the transition energy
$\varepsilon_{trans}$ can reach relatively small values in the two
energy-level crossing regions, i.e., the energy eigenvalues of the
doubly-occupied branch approach that of the two singly-occupied branches, and
the energy eigenvalues of the two singly-occupied branches approach that of
the unoccupied branch. Particularly, the switching time can reach a
minimum value when the sequential tunneling gap $\varepsilon_{se}$ is
approximately equal to zero, which corresponds to the two energy-level
crossing points, see Fig. 3 and the $\varepsilon_{\text{LUMO}}-V_{g}=1.0$ and
$\varepsilon_{\text{LUMO}}-V_{g}=-26.0$ cases in Fig. 4.

Next, we discuss the influence of the bias voltage on the spin-state switching
processes. For the given LUMO's energy $\varepsilon_{\text{LUMO}}-eV_{g}$ and
spin polarizations of the source and drain electrodes, with increasing the
applied bias voltage, new electron tunneling channels can be opened, which
further increase the electron tunneling processes associated with the
spin-state switching. The switching processes are thus enhanced rapidly, and
the corresponding switching time then decreases, see Fig. 5. However, When the
applied bias voltage is larger than a threshold bias voltage that depending on
the SMM's internal level structure, all tunneling channels are opened, and
then the switching time approaches a saturated value. see the $V_{b}=\pm10$
and $V_{b}=\pm12$ cases in Fig. 5. Therefore, the switching time decreases
with increasing the applied bias voltage at first and approaches a minimum
value finally at the so-called saturated bias voltage.

Furthermore, the interchange between the two spin-state of the $\left\langle
S_{z}\right\rangle _{\max}$ (i.e., the information bits "$1$") and the
$\left\langle S_{z}\right\rangle _{\min}$ (i.e., the information bits "$0$")
is an important aspect of the error correction of the spin states of stored information.
Figure 6 displays the transition from the spin-state of the $\left\langle
S_{z}\right\rangle _{\max}$ to the spin-state of the $\left\langle S_{z}\right\rangle
_{\min}$ or in the opposite way from the spin-state of the $\left\langle
S_{z}\right\rangle _{\min}$ to the spin-state of the $\left\langle S_{z}\right\rangle
_{\max}$. Here, the initial spin-state of the $\left\langle S_{z}\right\rangle
_{\max}$ and the $\left\langle S_{z}\right\rangle _{\min}$ are chosen as the
steady spin-state in Fig. 4. This operation can be simply carried out by
reversing the direction of the writing bias voltage. Therefore, in the
$P_{L}=-P_{R}$ case, the direction of the bias voltage can realize the fast
switching from a given SMM's spin-state to the spin-state of the $\left\langle
S_{z}\right\rangle _{\max}$ or the $\left\langle S_{z}\right\rangle _{\min}$,
which is independent of the initial spin-state. Here, it is noted to that in
the $P_{L}\neq-P_{R}$ case the values of the spin-state of the $\left\langle
S_{z}\right\rangle _{\max}$ and the $\left\langle S_{z}\right\rangle _{\min}$
have a very small difference for the different applied bias voltages. The
difference of these values is about the order of $10^{-2}$ for the SMM's
parameters considered here.

\subsection{Readout of spin states of stored information}

Another important processing is how to read out the two spin-state of the
$\left\langle S_{z}\right\rangle _{\max}$ and the $\left\langle S_{z}\right\rangle
_{\min}$. In order to reduce the influence of the readout processes on the two
spin-state of stored information, it requires a small bias voltage and a
short readout time. Here, the magnitude of the readout bias voltage is chosen
as $V_{b}=0$, namely, the writing bias voltage of the spin-state of the
$\left\langle S_{z}\right\rangle _{\max}$ or the $\left\langle S_{z}%
\right\rangle _{\min}$ is turned off. In this situation, the tunneling
processes of conduction electrons are strongly suppressed, and then the
occupation probabilities of the spin-state of the $\left\langle
S_{z}\right\rangle _{\min}$ and the $\left\langle S_{z}\right\rangle _{\max}$
will undergo a nonequilibrium physical process before they relax to a
quasi-equilibrium state. Particularly, the time of the nonequilibrium physical
processes is of the order of nanoseconds, see Figs. 7 and 8. We consider the
$V_{b}\left(  \text{W}\right)  =\pm6.0$ case here.

In the initial spin-state of the $\left\langle S_{z}\right\rangle _{\max}$ (or
the $\left\langle S_{z}\right\rangle _{\min}$), $\left\vert
1,S+1/2\right\rangle $ (or $\left\vert 1,-S-1/2\right\rangle $) has the
maximum occupation probability, while $\left\vert 2,S\right\rangle $ (or
$\left\vert 2,-S\right\rangle $) and $\left\vert 0,S\right\rangle $ (or
$\left\vert 0,-S\right\rangle $) have a very small occupation probabilities.
When the spin-state of the $\left\langle S_{z}\right\rangle _{\max}$ (or the
$\left\langle S_{z}\right\rangle _{\min}$) relaxes to a quasi-equilibrium
state, in the cases of the unoccupied, singly-occupied and doubly-occupied
ground eigenstates, the eigenstates $\left\vert 0,S\right\rangle $ (or
$\left\vert 0,-S\right\rangle $), $\left\vert 1,S+1/2\right\rangle $ (or
$\left\vert 1,-S-1/2\right\rangle $) and $\left\vert 2,S\right\rangle $ (or
$\left\vert 2,-S\right\rangle $) have the maximum occupation probabilities,
respectively. Therefore, the relaxation of the occupation probabilities of the
spin-state of the $\left\langle S_{z}\right\rangle _{\max}$ and the
$\left\langle S_{z}\right\rangle _{\min}$ can induce a current, e.g.,
$\left\vert 1,S+1/2\right\rangle \overset{\text{out: }\uparrow}%
{\longrightarrow}\left\vert 0,S\right\rangle $, $\left\vert
1,-S-1/2\right\rangle \overset{\text{out: }\downarrow}{\longrightarrow
}\left\vert 0,-S\right\rangle $, $\left\vert 1,S+1/2\right\rangle
\overset{\text{into: }\downarrow}{\longrightarrow}\left\vert 2,S\right\rangle
$ and $\left\vert 1,-S-1/2\right\rangle \overset{\text{into: }\uparrow
}{\longrightarrow}\left\vert 2,-S\right\rangle $, although that decreases to
zero very rapidly at a few nanoseconds, see Figs. 7 and 8.

In the case of the unoccupied ground eigenstates, in the nonequilibrium
relaxation processes, the spin-down electron tunnels out the SMM and into the
right-electrode easily due to $\Gamma_{R}^{\downarrow}\gg\Gamma_{R}^{\uparrow
}$, which corresponds to the transition $\left\vert 1,-S-1/2\right\rangle
\overset{\text{out: }\downarrow}{\longrightarrow}\left\vert 0,-S\right\rangle
$. Then, the time derivative of the transport current for the spin-state of
the $\left\langle S_{z}\right\rangle _{\min}$ shows much more remarkable
variation over time than that for the spin-state of the $\left\langle
S_{z}\right\rangle _{\max}$, see Figs. 7(a) and 7(b).
Whereas in the case of the doubly-occupied ground eigenstates,
the spin-down electron tunnels into the SMM from the right-electrode
easily due to $\Gamma_{R}^{\downarrow}\gg\Gamma_{R}^{\uparrow}$, which
corresponds to the transition $\left\vert 1,S+1/2\right\rangle \overset
{\text{into: }\downarrow}{\longrightarrow}\left\vert 2,S\right\rangle $, then,
the time derivative of the transport current for the spin-state of the
$\left\langle S_{z}\right\rangle _{\max}$ shows much more obvious variation
over time than that for the spin state of the $\left\langle S_{z}\right\rangle
_{\min}$, see Figs. 8(c) and 8(d).

As for the case of the singly-occupied ground eigenstates and the electron
tunneling processes being dominated by the transitions between the
singly-occupied and unoccupied eigenstates, when the spin-state of the
$\left\langle S_{z}\right\rangle _{\max}$ (or the $\left\langle S_{z}%
\right\rangle _{\min}$) relaxes to a quasi-equilibrium state, in the case of
the value of $\varepsilon_{\left\vert 0,S\right\rangle }$ being slightly
larger than that of $\varepsilon_{\left\vert 1,S+1/2\right\rangle }$, e.g.,
$\varepsilon_{\text{LUMO}}-V_{g}=0.0$, the occupation probability of the
$\left\vert 1,S+1/2\right\rangle $ (or $\left\vert 1,-S-1/2\right\rangle $)
decreases while that of the $\left\vert 0,S\right\rangle $ (or $\left\vert
0,-S\right\rangle $) increases. In this situation, since $\Gamma
_{R}^{\downarrow}\gg\Gamma_{R}^{\uparrow}$ the transition $\left\vert
1,-S-1/2\right\rangle \overset{\text{out: }\downarrow}{\longrightarrow
}\left\vert 0,-S\right\rangle $ occurs more quickly than $\left\vert
1,S+1/2\right\rangle \overset{\text{out: }\uparrow}{\longrightarrow}\left\vert
0,S\right\rangle $, then, the time derivative of the transport current for the
spin-state of the $\left\langle S_{z}\right\rangle _{\min}$ shows relatively
obvious variation over time than that for the spin-state of the $\left\langle
S_{z}\right\rangle _{\max}$, see the $\varepsilon_{\text{LUMO}}-V_{g}=0.0$
case in Figs. 7(c) and 7(d). While in the case of the value of $\varepsilon_{\left\vert
0,S\right\rangle }$ being much larger than that of $\varepsilon_{\left\vert
1,S+1/2\right\rangle }$, e.g., $\varepsilon_{\text{LUMO}}-V_{g}=-2.0$, the
occupation probability of the $\left\vert 1,S+1/2\right\rangle $ (or
$\left\vert 1,-S-1/2\right\rangle $) further increases while that of the
$\left\vert 0,S\right\rangle $ (or $\left\vert 0,-S\right\rangle $) decreases.
In this situation, although the transition $\left\vert 0,-S\right\rangle
\overset{\text{into: }\downarrow}{\longrightarrow}\left\vert
1,-S-1/2\right\rangle $ occurs more quickly than $\left\vert
0,S\right\rangle \overset{\text{into: }\uparrow}{\longrightarrow}\left\vert
1,S+1/2\right\rangle $ due to $\Gamma_{R}^{\downarrow}\gg\Gamma_{R}^{\uparrow
}$, the magnitude of the occupation probabilities of the $\left\vert
0,-S\right\rangle $ and $\left\vert 0,S\right\rangle $ is very small.
Therefore, the time derivative of the transport current for the spin-state of
the $\left\langle S_{z}\right\rangle _{\min}$ does not show an obvious over
time than that for the spin-state of the $\left\langle S_{z}\right\rangle
_{\max}$, see the $\varepsilon_{\text{LUMO}}-V_{g}=-1.0$ and $\varepsilon
_{\text{LUMO}}-V_{g}=-2.0$ cases in Figs. 7(c) and 7(d).

Whereas in the case of the singly-occupied ground eigenstates and the electron
tunneling processes being dominated by the transitions between the
doubly-occupied and singly-occupied eigenstates, for the spin-state of the
$\left\langle S_{z}\right\rangle _{\max}$ (or the $\left\langle S_{z}%
\right\rangle _{\min}$), the occupation probability of the $\left\vert
2,S\right\rangle $ (or $\left\vert 2,-S\right\rangle $) is much larger than
that of the $\left\vert 0,S\right\rangle $ (or $\left\vert 2,-S\right\rangle
$). In the nonequilibrium relaxation processes, when the value of
$\varepsilon_{\left\vert 2,S\right\rangle }$ is much larger than that of
$\varepsilon_{\left\vert 1,S+1/2\right\rangle }$, e.g., $\varepsilon
_{\text{LUMO}}-V_{g}=-23.0$ and $\varepsilon_{\text{LUMO}}-V_{g}=-24.0$, the
occupation probability of the $\left\vert 1,S+1/2\right\rangle $ (or
$\left\vert 1,-S-1/2\right\rangle $) further increases while that of the
$\left\vert 2,S\right\rangle $ (or $\left\vert 2,-S\right\rangle $) and
$\left\vert 0,S\right\rangle $ (or $\left\vert 0,-S\right\rangle $) decrease.
Although the spin-down electron tunnels out SMM and into the the right-electrode
easily due to $\Gamma_{R}^{\downarrow}\gg\Gamma_{R}^{\uparrow}$, which
corresponds to the transition $\left\vert 2,S\right\rangle \overset{\text{out:
}\downarrow}{\longrightarrow}\left\vert 1,S+1/2\right\rangle $, the magnitude
of the occupation probability of the $\left\vert 2,S\right\rangle $ is very
small here. Therefore, the time derivative of the transport current for the
spin-state of the $\left\langle S_{z}\right\rangle _{\max}$ shows much less
obvious variation over time than that for the spin-state of the $\left\langle
S_{z}\right\rangle _{\min}$, see the $\varepsilon_{\text{LUMO}}-V_{g}=-23.0$
and $\varepsilon_{\text{LUMO}}-V_{g}=-24.0$ cases in Figs. 8(a) and 8(b). While the
value of $\varepsilon_{\left\vert 2,S\right\rangle }$ is slightly larger than
that of $\varepsilon_{\left\vert 1,S+1/2\right\rangle }$, e.g., $\varepsilon
_{\text{LUMO}}-V_{g}=-25.0$, the occupation probability of the $\left\vert
1,S+1/2\right\rangle $ (or $\left\vert 1,-S-1/2\right\rangle $) in the
nonequilibrium relaxation processes decreases while that of the $\left\vert
2,S\right\rangle $ (or $\left\vert 2,-S\right\rangle $) increases. In this
situation, since $\Gamma_{R}^{\downarrow}\gg\Gamma_{R}^{\uparrow}$ the
transition $\left\vert 1,S+1/2\right\rangle \overset{\text{into: }\downarrow
}{\longrightarrow}\left\vert 2,S\right\rangle $ occurs more quickly than that
$\left\vert 1,-S-1/2\right\rangle \overset{\text{into: }\uparrow
}{\longrightarrow}\left\vert 2,-S\right\rangle $, the time derivative of the
transport current for the spin-state of the $\left\langle S_{z}\right\rangle
_{\max}$ shows relatively obvious variation over time than that for the
spin-state of the $\left\langle S_{z}\right\rangle _{\min}$, see the
$\varepsilon_{\text{LUMO}}-V_{g}=-25.0$ case in Figs. 8(a) and 8(b).

Consequently, in the cases of the unoccupied and doubly-occupied ground
eigenstates, the time derivative of the transport current $dI/dt$ as a
function of time suggests a possible quantity for the readout of the two spin-state
of the $\left\langle S_{z}\right\rangle _{\max}$ and
the $\left\langle S_{z}\right\rangle _{\min}$, see Figs. 7(a) and 7(b), and Figs. 8(c) and 8(d). However,
in the case of the singly-occupied ground eigenstates, the time derivative of
the transport current for the spin-state of the $\left\langle S_{z}%
\right\rangle _{\max}$ does not show an obvious variation over time than that
for the spin-state of the $\left\langle S_{z}\right\rangle _{\min}$ although
the switching process might have a relatively small switching time, see
Fig. 4, and Figs. 7(c) and 7(d), and Figs. 8(a) and 8(b).

\subsection{Dependence of the switching time and the readout time on the
SMM-electrode coupling}

To illustrate that the influences of the asymmetry of the SMM-electrode
coupling and the corresponding coupling strengths on the switching time and
the readout time, we consider here the $\varepsilon_{\text{LUMO}}-eV_{g}=2.0$
case. For the given writing bias voltage and spin polarizations of the source
and drain electrodes, the switching time is determined by the number of
conduction electrons per unit time tunneling into the SMM from the source electrode,
namely, $\Gamma_{L}$, and the number of conduction electrons per unit time tunneling out the SMM and
into the drain electrode, namely, $\Gamma_{R}$. Therefore, the SMM-electrode
coupling and the corresponding coupling strengths have a strong influence on
the switching time, see Fig. 9. Furthermore, although the magnitudes of the
$dI/dt$ of the spin-state of stored information over time in the readout
process depend on the the asymmetry of the SMM-electrode coupling and the
corresponding coupling strengths, the relaxation time is mainly determined by
the SMM's internal level structure. Consequently, the asymmetry of the
SMM-electrode coupling and the corresponding coupling strengths have a slight
influence upon the readout time, which is still of the order of nanosecond,
see Fig. 10, and can also be understood in terms of the above mentioned
relaxation of the SMM's eigenstates.

\section{CONCLUSIONS}

We have analyzed the feasibility of the SMM-based memory device. Here, the
SMM's spin-state of the $\left\langle S_{z}\right\rangle _{\max}$ and that of
the $\left\langle S_{z}\right\rangle _{\min}$ are recognized as the
information bits \textquotedblleft$1$" and \textquotedblleft$0$",
respectively. We have demonstrated that the switching between the SMM's two
spin states of stored information (i.e., \textquotedblleft$1$" and
\textquotedblleft$0$") can be carried out by reversing the direction of the
applied spin-polarized current. This feature may be useful for the error
correction of stored information. The switching time, which depends on both
the sequential tunneling gap $\varepsilon_{se}$ and the spin-selection-rule
allowed transition-energy $\varepsilon_{trans}$, can be tuned by the gate
voltage. In particular, when the writing bias voltage is turned off,
in the cases of the unoccupied and doubly-occupied
ground eigenstates, the time derivative of the transport current can be used
to read out the SMM's two spin-state of the $\left\langle S_{z}\right\rangle
_{\max}$ and the $\left\langle S_{z}\right\rangle _{\min}$. Moreover,
the tunneling strength of and the asymmetry of SMM-electrode coupling have a
strong influence on the switching time, whereas that have a slight influence on
the readout time. These results provide fundamental insight into the
electrical controllable manipulation and detection of the SMM's two spin-state of
stored information, establishing the basis for the application of the SMM in
memory device at the single-molecule level.

\section{ACKNOWLEDGMENTS}

This work was supported by the Shanxi Natural Science Foundation of China
under Grant No. 201601D011015, Program for the Outstanding Innovative Teams of
Higher Learning Institutions of Shanxi, National Key R\&D Program of China
under Grant No. 2016YFA0301500, NSFC under Grants Nos. 11204203, 11434015,
61227902, 61835013, KZ201610005011, SPRPCAS under Grants Nos. XDB01020300, XDB21030300.

\newpage

\begin{figure}[t]
\centerline{\includegraphics[height=10cm,width=16cm]{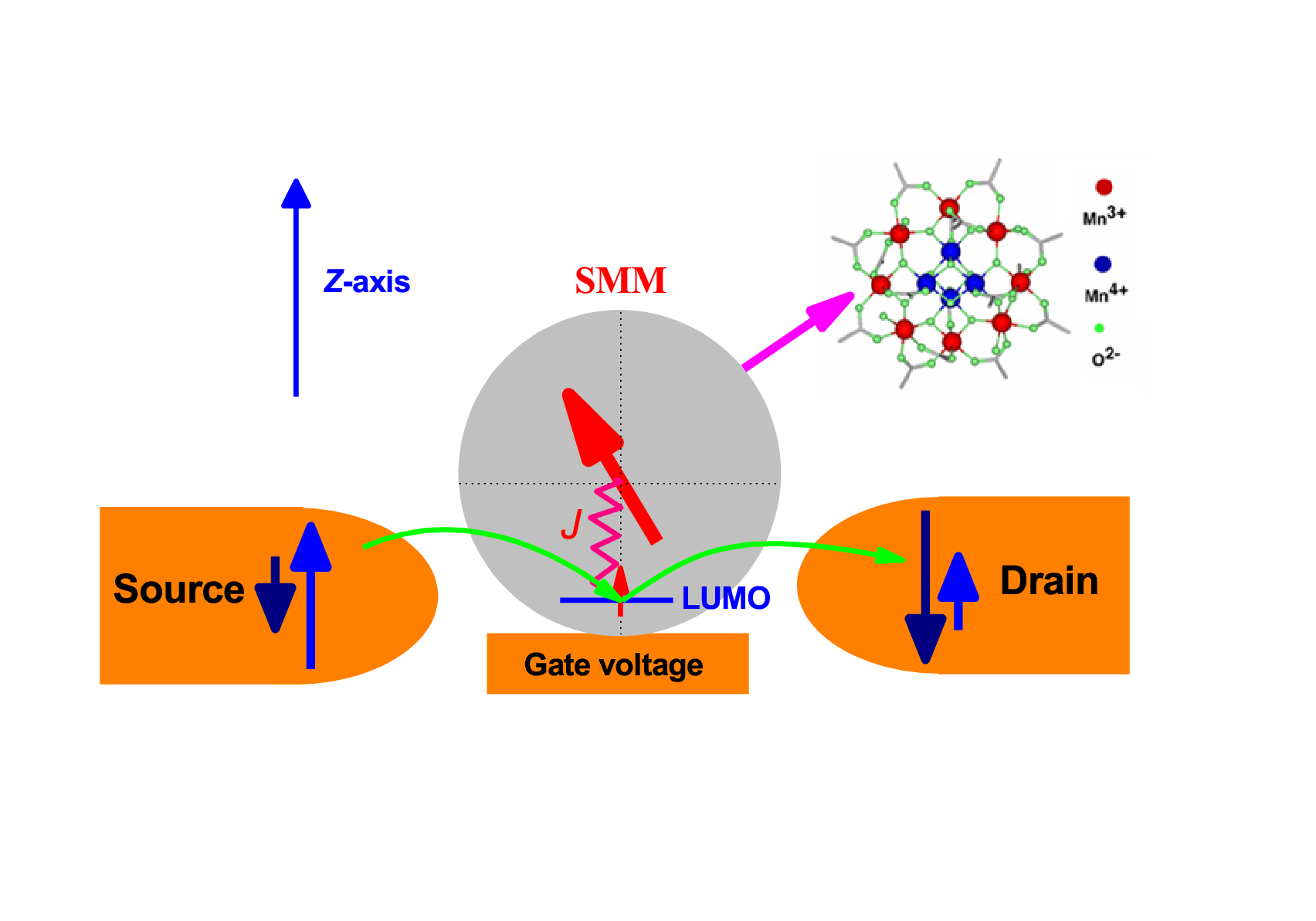}}\caption{(Color online) Schematic of a single-molecule memory device
consisting of a SMM Mn$_{12}$ weakly coupled to the two ferromagnetic
electrodes (leads). The SMM's two spin states of the $\left\langle
S_{z}\right\rangle _{\max}$ and the $\left\langle S_{z}\right\rangle _{\min}$
are defined as the two states of the binary code, the spin-up ($1$) and
spin-down ($0$), respectively. Here, $\left\langle S_{z}\right\rangle $,
namely, the average value of the $Z$-component of the total spin of the SMM and the conduction-electron is
calculated as $\left\langle S_{z}\right\rangle ={\displaystyle\sum\nolimits
_{n,m}}mP_{\left\vert n,m\right\rangle }$, where $P_{\left\vert
n,m\right\rangle }$ is the occupation probability of the SMM's eigenstate
$\left\vert n,m\right\rangle $ with $n$ ($n=0,1,2$) being the electron
number and $m$ the magnetic quantum-number of the total spin.}%
\label{fig1}%
\end{figure}

\newpage

\begin{figure}[t]
\centerline{\includegraphics[height=10cm,width=16cm]{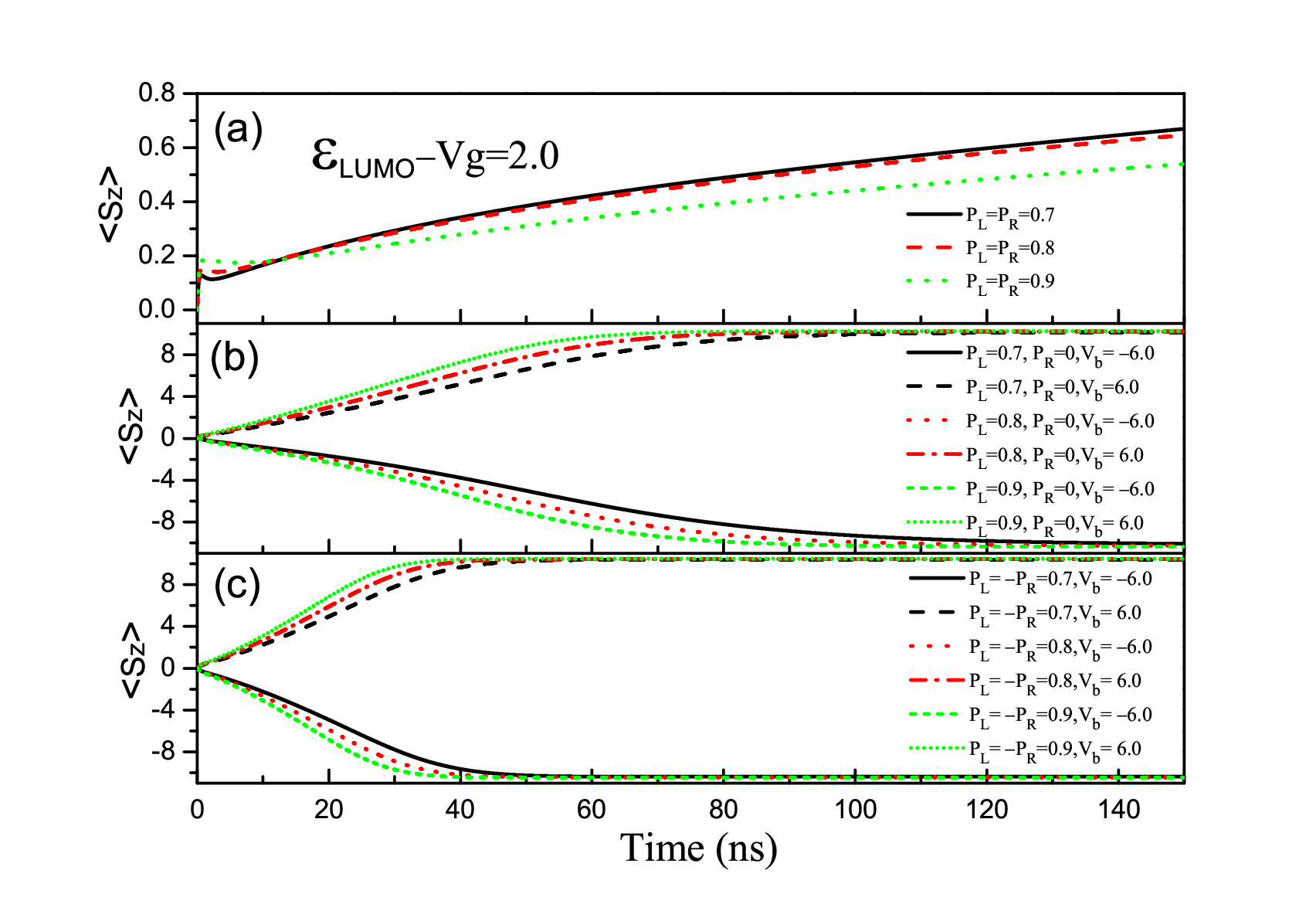}}\caption{(Color
online) The average value of the $\left\langle
S_{z}\right\rangle $ as a function of time for different values of $P_{L}$ and
$P_{R}$ with $\varepsilon_{d}-eV_{g}=2.0$ and $V_{b}=\pm6.0$. Here, we
consider the three different configurations (a), $P_{L}=P_{R}=p$, (b),
$P_{L}=p$ and $P_{R}=0$ and (c), $P_{L}=-P_{R}=p$. The SMM's parameters:
$S=10$, $U=25$, $J=0.2$, $K_{1}=0.06$, $\Gamma_{L}=\Gamma_{R}=0.002$ and
$k_{B}T=0.4$, where meV has been chosen as the unit of energy.}%
\label{fig2}%
\end{figure}

\newpage

\begin{figure}[t]
\centerline{\includegraphics[height=10cm,width=16cm]{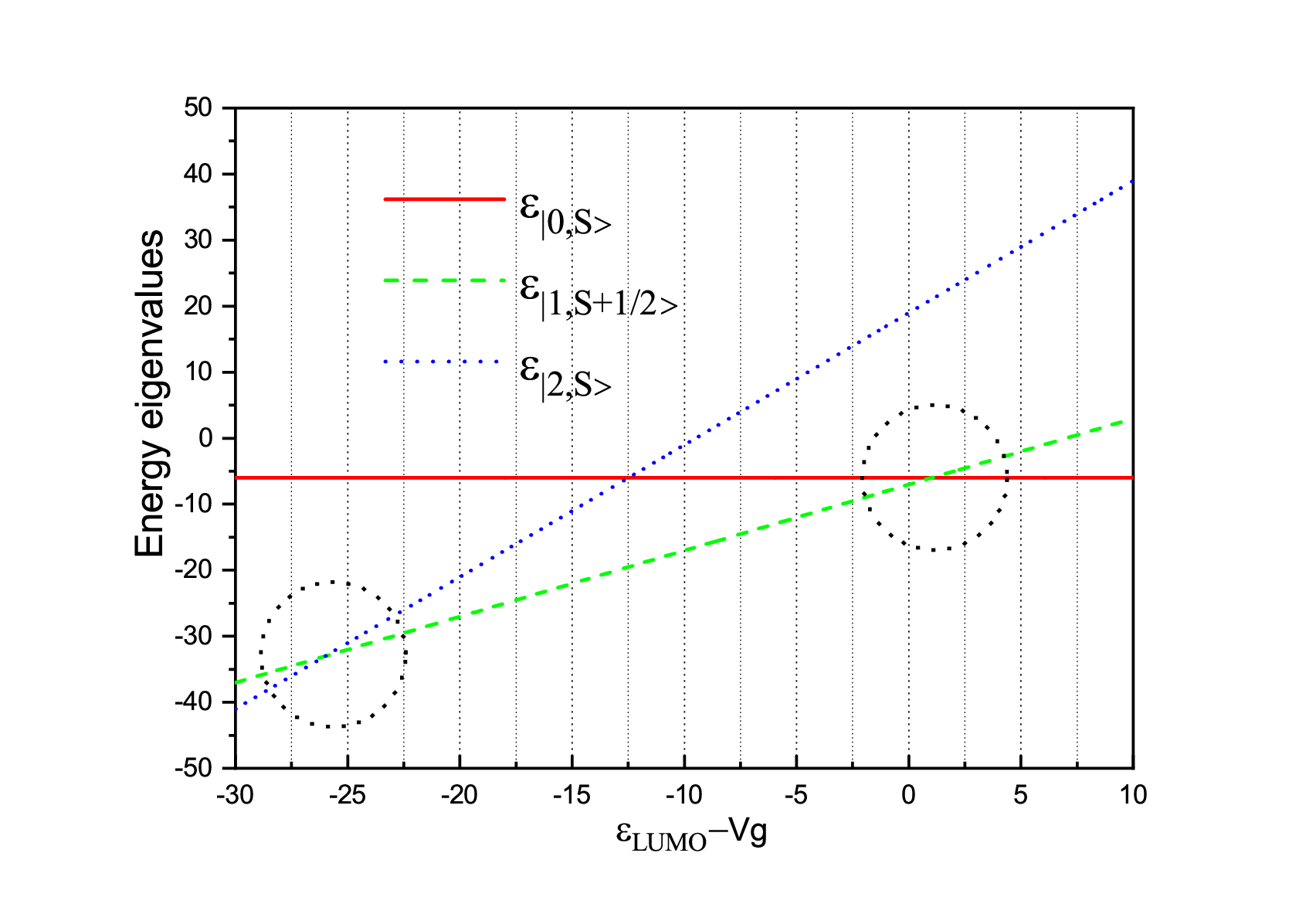}}\caption{(Color
online) The minimum energy eigenvalues of the unoccupied, singly-occupied
and doubly-occupied eigenstates as a function of $\varepsilon
_{\text{LUMO}}-eV_{g}$. The SMM's other parameters are the same as in Fig. 2.}%
\label{fig3}%
\end{figure}

\newpage

\begin{figure}[t]
\centerline{\includegraphics[height=10cm,width=16cm]{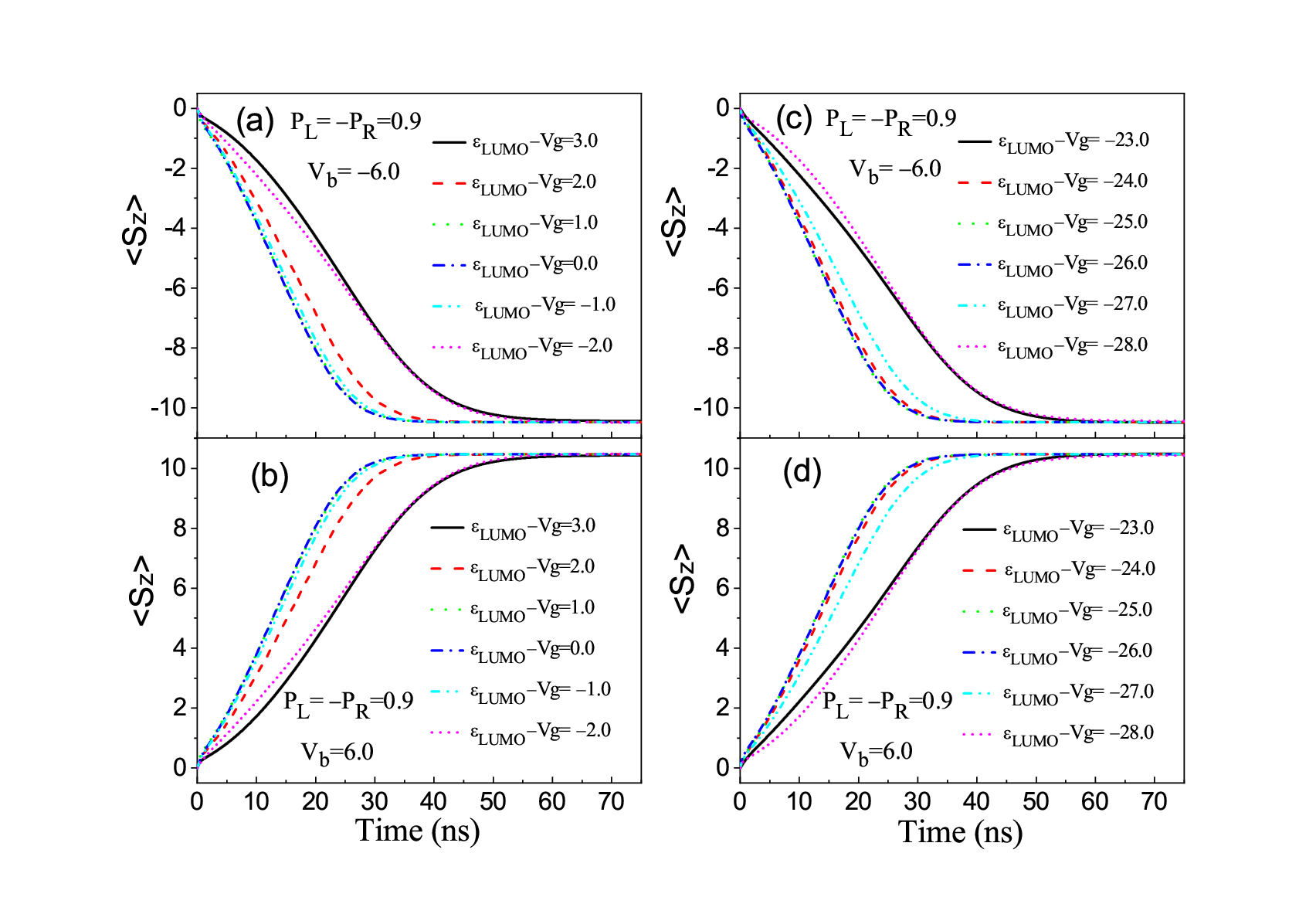}}\caption{(Color
online) The average value
of the $\left\langle S_{z}\right\rangle $ as a function of time for different
values of $\varepsilon_{\text{LUMO}}-eV_{g}$ with $P_{L}=-P_{R}=0.9$ and $\Gamma_{L}=\Gamma_{R}=0.002$. (a) and
(c), $V_{b}=-6.0$, (b) and (d), $V_{b}=6.0$. The SMM's other parameters are
the same as in Fig. 2.}%
\label{fig4}%
\end{figure}

\newpage

\begin{figure}[t]
\centerline{\includegraphics[height=10cm,width=16cm]{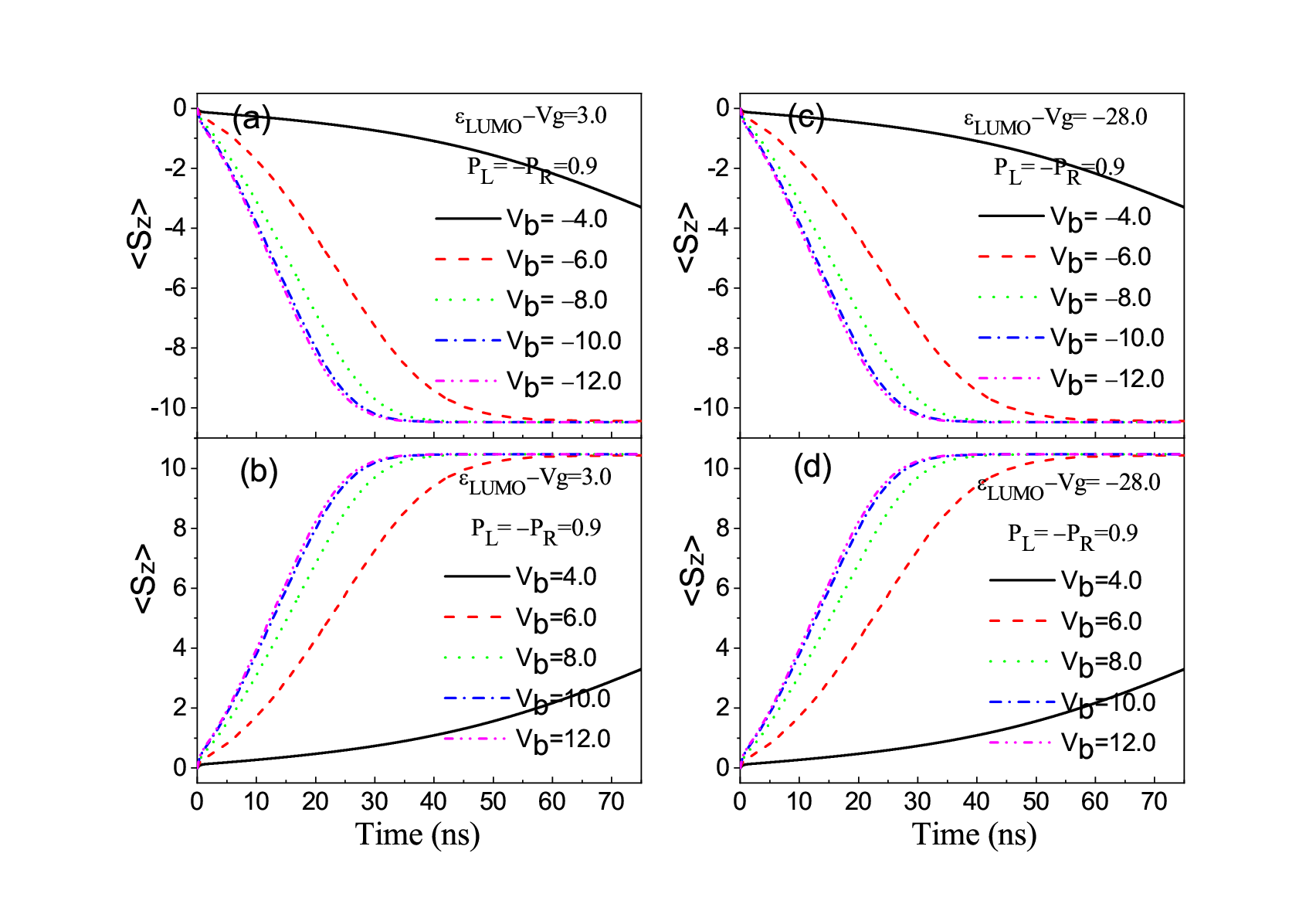}}\caption{(Color
online) The average value of $\left\langle
S_{z}\right\rangle $ as a function of time for different values of $V_{b}$ with
$P_{L}=-P_{R}=0.9$ and $\Gamma_{L}=\Gamma_{R}=0.002$. (a) and (b), $\varepsilon_{\text{LUMO}}-eV_{g}=3.0$, (c)
and (d), $\varepsilon_{\text{LUMO}}-eV_{g}=-28.0$. The SMM's other parameters
are the same as in Fig. 2.}%
\label{fig5}%
\end{figure}

\newpage

\begin{figure}[t]
\centerline{\includegraphics[height=10cm,width=16cm]{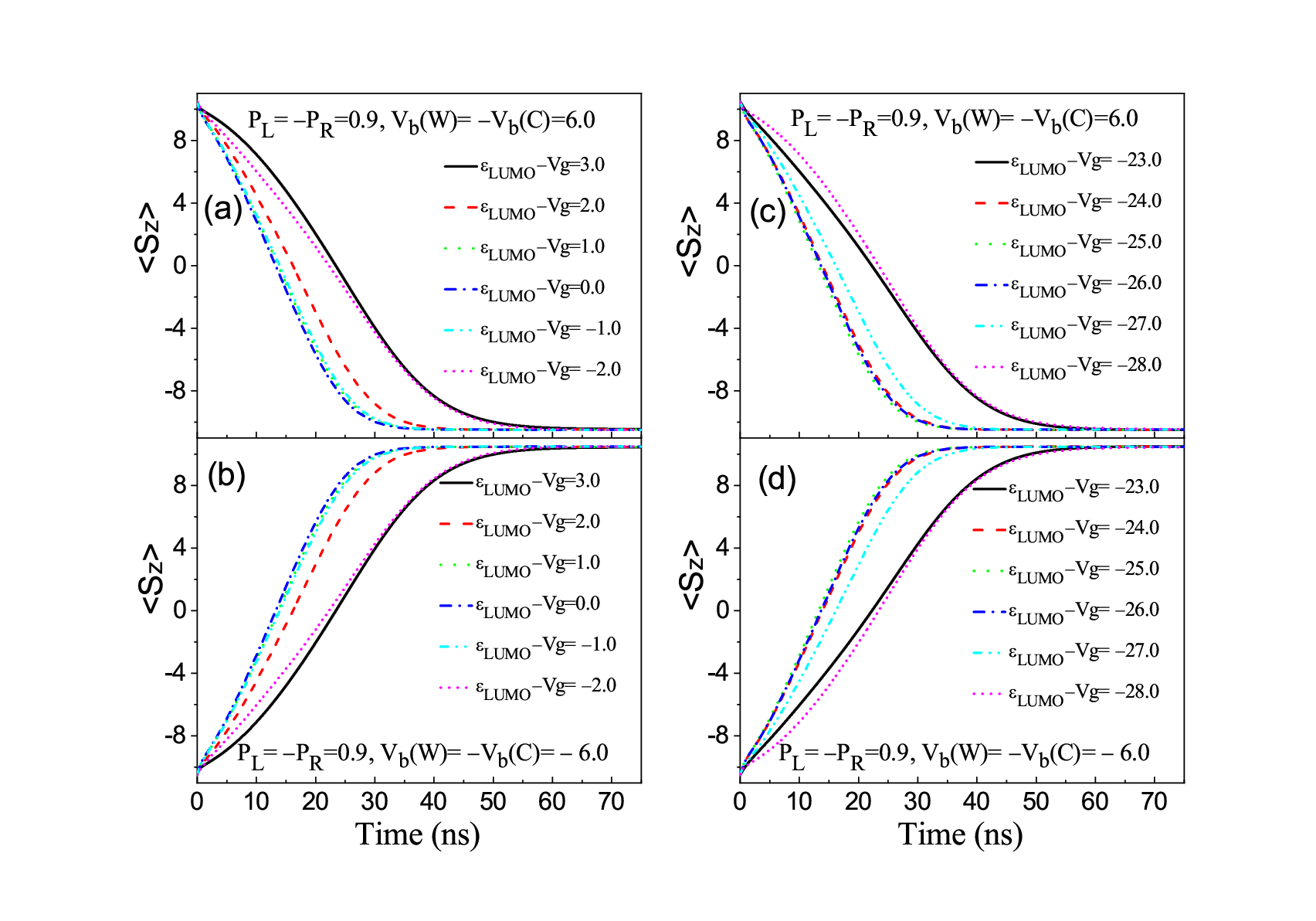}}\caption{(Color
online) The average value of the $\left\langle S_{z}\right\rangle $ as a
function of time for different values of $\varepsilon_{\text{LUMO}}-eV_{g}$ with
$P_{L}=-P_{R}=0.9$ and $\Gamma_{L}=\Gamma_{R}=0.002$. Here, the initial spin states of the $\left\langle
S_{z}\right\rangle _{\max}$ and the $\left\langle S_{z}\right\rangle _{\min}$
are chosen as the steady spin states in Fig. 4, and the magnitude of the
controlling bias voltage $V_{b}(\text{C})$ is chosen the same as that of the
writing one $V_{b}(\text{W})$. (a) and (c), $V_{b}(\text{W})=-V_{b}%
(\text{C})=6.0$, (b) and (d), $V_{b}(\text{W})=-V_{b}(\text{C})=-6.0$. The
SMM's other parameters are the same as in Fig. 2.}%
\label{fig6}%
\end{figure}

\newpage

\begin{figure}[t]
\centerline{\includegraphics[height=12cm,width=16cm]{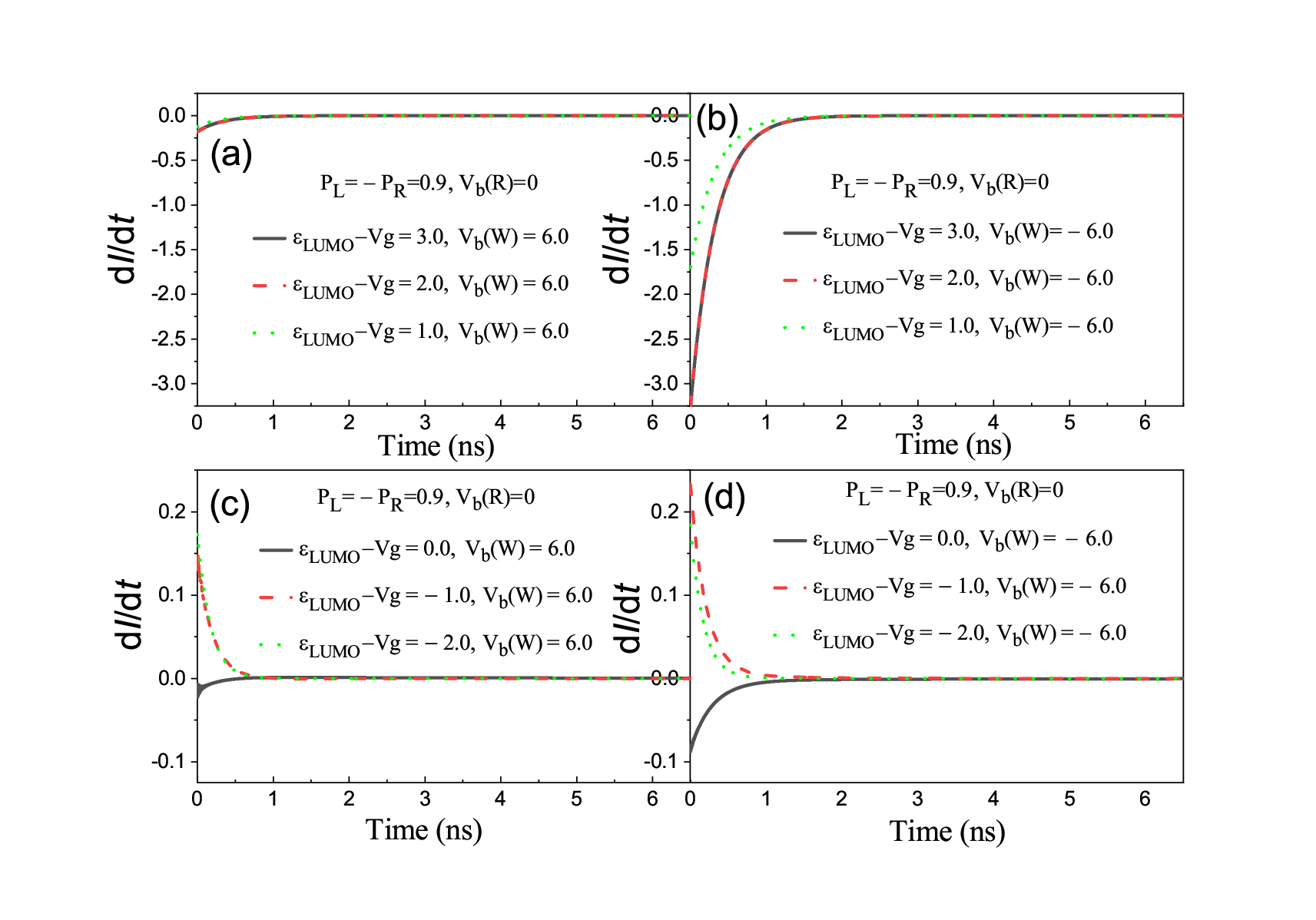}}\caption{(Color
online) The time derivative of the
transport current, namely, $dI/dt$, as a function of time for different spin
states of stored information with $P_{L}=-P_{R}=0.9$ and $\Gamma_{L}=\Gamma
_{R}=0.002$. (a) and (c), the spin-state of the $\left\langle S_{z}\right\rangle _{\max}$;
(b) and (d),  the spin-state of the $\left\langle
S_{z}\right\rangle _{\min}$. The SMM's other parameters are the same as in Fig. 2.}%
\label{fig7}%
\end{figure}

\newpage

\begin{figure}[t]
\centerline{\includegraphics[height=12cm,width=16cm]{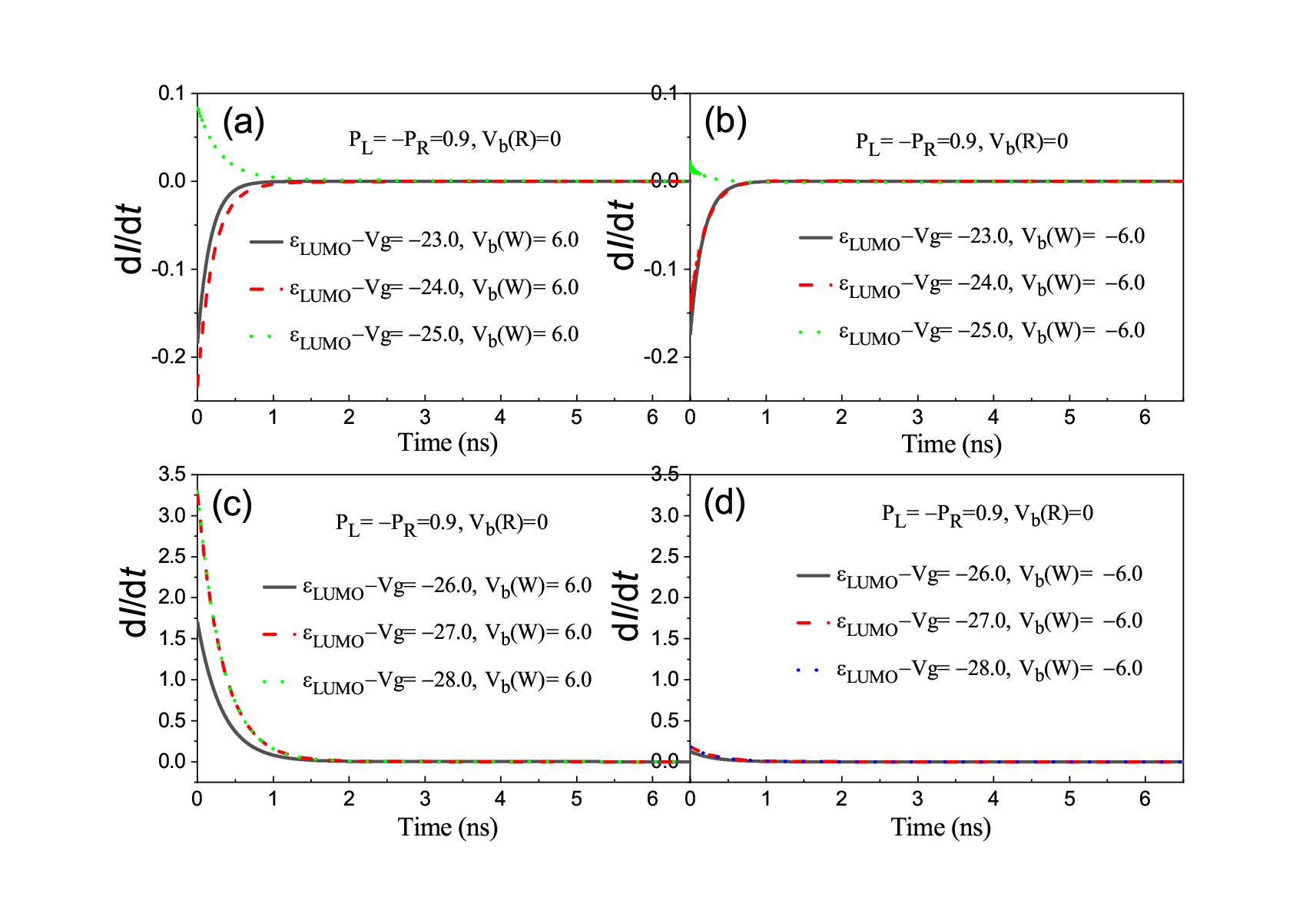}}\caption{(Color
online) The time derivative of the
transport current, namely, $dI/dt$, as a function of time for different spin
states of stored information with $P_{L}=-P_{R}=0.9$ and $\Gamma_{L}=\Gamma
_{R}=0.002$. (a) and (c), the spin-state of the $\left\langle S_{z}\right\rangle _{\max}$;
(b) and (d),  the spin-state of the $\left\langle
S_{z}\right\rangle _{\min}$. The SMM's other parameters are the same as in Fig. 2.}%
\label{fig8}%
\end{figure}

\newpage

\begin{figure}[t]
\centerline{\includegraphics[height=10cm,width=16cm]{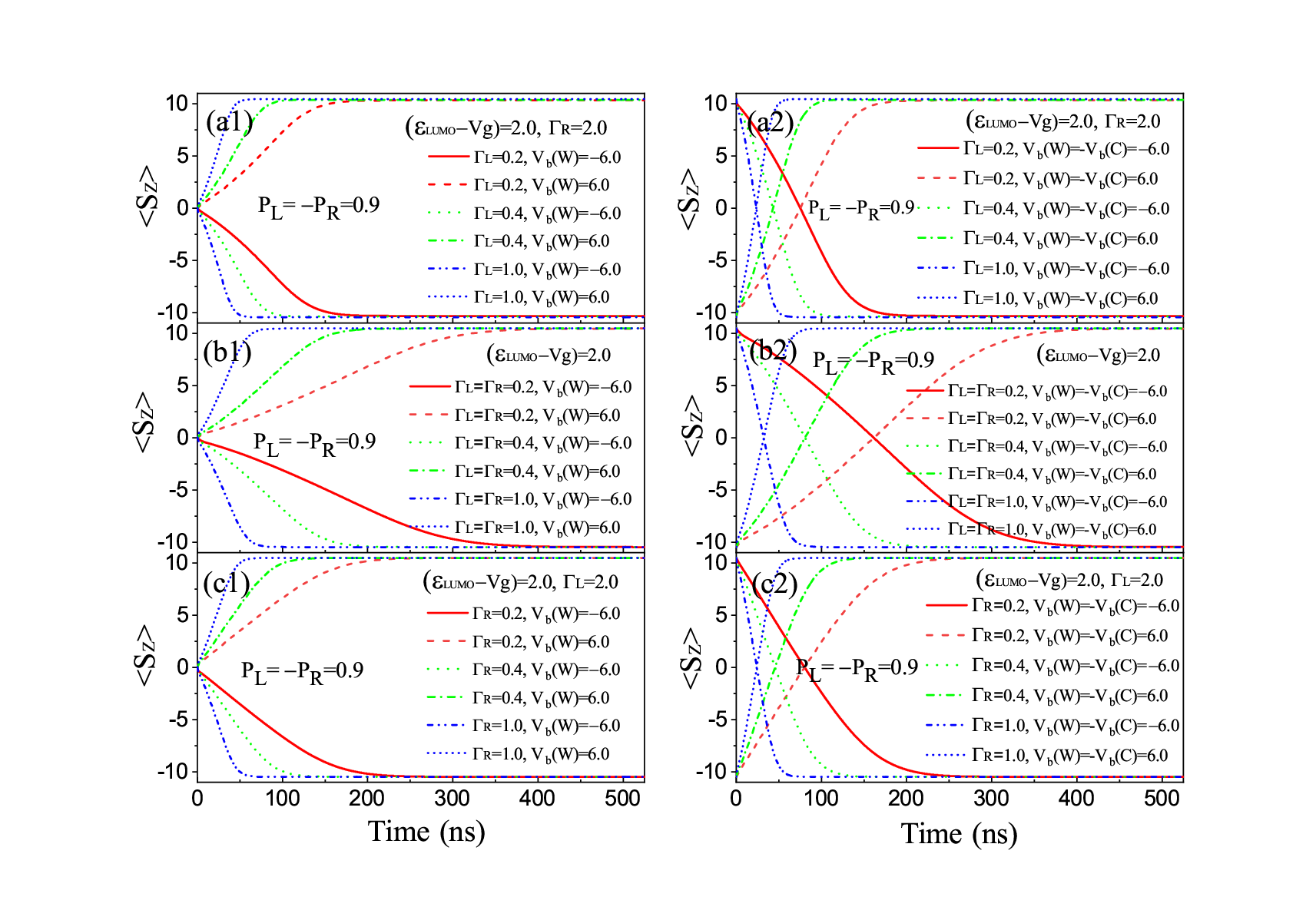}}\caption{(Color
online) The average value
of the $\left\langle S_{z}\right\rangle $ as a function of time for different
values of the tunneling strength of and the asymmetry of SMM-electrode coupling.
(a1), (b1) and (c1), $P_{L}=-P_{R}=0.9$, $\varepsilon_{\text{LUMO}}-eV_{g}=2.0$ and $V_{b}(\text{W})=\pm6.0$;
(a2), (b2) and (c2), $P_{L}=-P_{R}=0.9$, $\varepsilon_{\text{LUMO}}-eV_{g}=2.0$ and $V_{b}(\text{W})=-V_{b}(\text{C})=6.0$.
(a1) and (a2), $\Gamma_{L}<\Gamma_{R}$, (b1) and (b2), $\Gamma_{L}=\Gamma_{R}$,
(c1) and (c2), $\Gamma_{L}>\Gamma_{R}$. The SMM's other parameters are the same as in Fig. 2.}%
\label{fig9}%
\end{figure}

\newpage

\begin{figure}[t]
\centerline{\includegraphics[height=14cm,width=16cm]{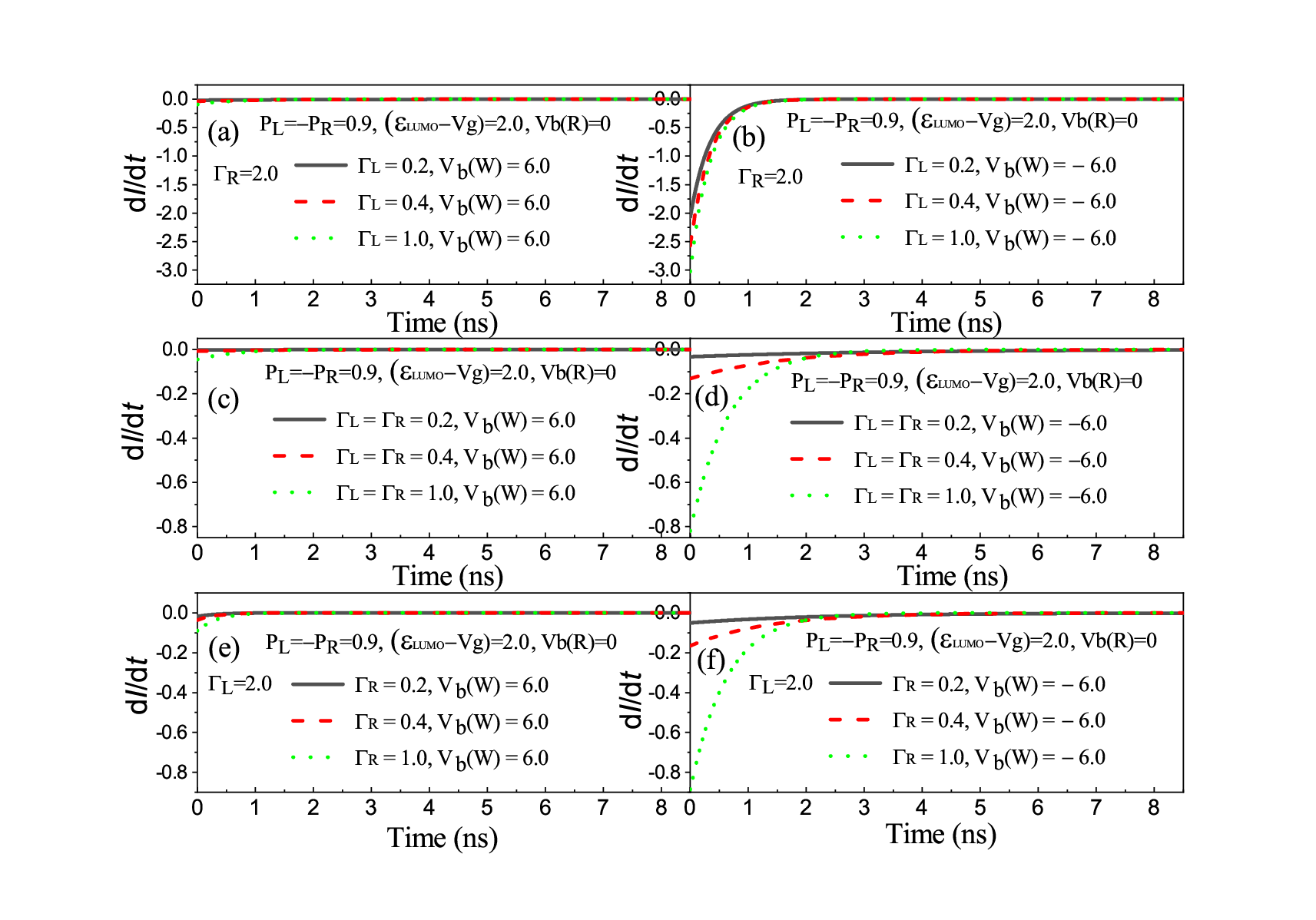}}\caption{(Color
online) The time derivative of the transport current, namely, $dI/dt$, as a function
of time for different spin states of stored information with $P_{L}=-P_{R}=0.9$, $\varepsilon
_{\text{LUMO}}-eV_{g}=2.0$ and $V_{b}(\text{W})=\pm6.0$. (a1), (b1) and (c1)
the spin-state of the $\left\langle S_{z}\right\rangle
_{\max}$; (a2), (b2) and (c2) the spin-state of the $\left\langle S_{z}\right\rangle _{\max}$.
(a1) and (a2), $\Gamma_{L}<\Gamma_{R}$, (b1) and (b2), $\Gamma_{L}=\Gamma_{R}$, (c1) and (c2),
$\Gamma_{L}>\Gamma_{R}$. The SMM's other parameters are the same as in Fig. 2.}%
\label{fig10}%
\end{figure}

\end{document}